\documentclass[dvips,10pt]{article}

\usepackage{a4wide,amsmath,amssymb,epsf,epsfig,amsthm,bm}

\title{Intersecting hypersurfaces, topological densities
and Lovelock Gravity}
\author{Elias Gravanis$^1$\footnote{E-mail:
eliasgravanis@netscape.net} and Steven Willison$^{1,2}$
\footnote{E-mail:
steve-at-cecs.cl}\\\\
{\it $^1$Department of Physics, Kings College, Strand, London WC2R
2LS, U.K.}
\\
{\it $^2$Centro de Estudios Cient\'{\i}ficos (CECS), Casilla 1469,
Valdivia, Chile.} }
\date{March 22, 2007}

\begin{document}

\maketitle

{\abstract Intersecting hypersurfaces in classical Lovelock gravity
are studied exploiting the description of the Lovelock Lagrangian as
a sum of dimensionally continued Euler densities. We wish to present
an interesting geometrical approach to the problem. The analysis
allows us to deal most efficiently with the division of space-time
into a honeycomb network of cells produced by an arbitrary
arrangement of membranes of matter. We write the gravitational
action as bulk terms plus integrals over each lower dimensional
intersection.

The spin connection is discontinuous at the shared boundaries of the
cells, which are spaces of various dimensionalities. That means that
at each intersection there are more than one spin connections.

We introduce a multi-parameter family of connections which
interpolate between the different connections at each intersection.
The parameters live naturally on a simplex. We can then write the
action including all the intersection terms in a simple way. The
Lagrangian of Lovelock gravity is generalized so as to live on the
simplices as well. Each intersection term of the action is then
obtained as an integral over an appropriate simplex.

Lovelock gravity and the associated topological (Euler) density are
used as an example of a more general formulation. In this example
one finds that singular sources up to a certain co-dimensionality
naturally carry matter without introducing conical or other
singularities in spacetime geometry.

}

\newpage

\section{Introduction}
In the light of the current trends in high energy physics, it is
widely supposed that space-time has dimensions higher than four.
In studying classical gravity, there are then other terms in the
gravitational action, yielding second order field equations, which
it is reasonable to consider. In $d$ dimensions we have the
general Lagrangian, first obtained by Lovelock~\cite{Lovelock}. We
use the vielbein formulation~\cite{Eguchi-80,Zumino-86}:
\begin{gather}\label{Shovelock}
{\cal L} =\sum _{n=0}^{[d/2]} \alpha_n f(\Omega^{\wedge n}\wedge
E^{\wedge d-2n}),
\\\nonumber
f(\Omega^{\wedge n}\wedge E^{\wedge d-2n}) = \Omega^{a_1a_2}
\wedge\cdot\cdot\cdot\wedge\Omega^{a_{2n-1}a_{2n}} \wedge
E^{a_{2n+1}}\wedge\cdot\cdot\cdot\wedge
E^{a_d}\epsilon_{a_1...a_{d}}.
\end{gather}
Above, $E^a$ are the vielbein frames and $\Omega^{ab}$ is the
curvature two-form:
\begin{gather*}
\Omega^{ab} = d\omega^{ab}+\frac{1}{2}[\omega,\omega]^{ab}
=d\omega^{ab}+\omega^a_c \wedge \omega^{cb}.
\end{gather*}
The spin connection is $\omega$. The Lie bracket for a $p$-form $A$
and $q$-form $B$ is $[A,B]^{ab} = A^a_c \wedge B^{cb}-(-1)^{pq}
B^a_c \wedge A^{cb}$. The totally antisymmetric tensor is normalized
to $\epsilon_{01\dots d} =+1$. $[d/2]$ is the highest integer less
than $d/2$. There are $[d/2]$ coefficients $\alpha_n$.

The first term in (\ref{Shovelock}) is the cosmological constant.
The second term is the Einstein-Hilbert Lagrangian. In the
familiar four dimensions it reads:
\begin{equation*}
{\cal L}_{EH}=\alpha_1 \Omega^{ab} \wedge E^c \wedge E^d
\epsilon_{abcd} = 2\alpha_1 R \sqrt{-g}
\end{equation*}
and the coefficient  in our conventions is related to Newton's
constant by $\alpha_1=(8\pi G)^{-1}$. In dimensions higher than
four, there are other terms in (\ref{Shovelock}) which are
corrections to the Einstein theory. Each term is a polynomial of
order $n$ in the curvature. These were studied in the late 1980's
when it was realised that they were related to strings and were
ghost free in a flat background~\cite{Zumino-86,Zwiebach}. The
relation to strings was further
discussed~\cite{Deser:1986xr,Metsaev:1986yb,Wiltshire:1988uq}, exact
solutions were obtained and black hole spacetimes were analyzed in
various
dimensions~\cite{Boulware:1985wk,Wheeler:1985nh,Wiltshire:1985us,
Myers:1986un,Gibbons:1987ps,Callan:1988hs,
Myers:1988ze,Jacobson:1993xs,Deruelle:1989fj}. Recently the special
properties of the theory have been studied
 motivated by braneworld models~\cite{Charmousis:2002rc,Deruelle},
higher dimensional black holes and also Chern-Simons
Gravity~\cite{Zanelli}. To us, the important property of the
Lagrangian (\ref{Shovelock}) is its affinity to topological
densities. It is the fact that will enable us to deal with the
problem of intersecting hypersurfaces filled with matter in this
theory with ease and generality. It will also allow us to write the
solution in an elegant form.

In $d=2n$, the term $f(\Omega^{\wedge n})$ is proportional to the
Euler density and is locally a total derivative. For example if we
write the Einstein-Hilbert term in two dimensions it reads
\begin{equation*}\label{}
\alpha_1  \Omega^{ab} \epsilon_{ab}
\end{equation*}
(the coupling $\alpha_1$ has dimension zero). Integrated over a
closed manifold $M$ it just gives the Euler number of the manifold.
No local information can be obtained from this. The analogous
quantity in arbitrary even dimension is
\begin{equation*}\label{}
\alpha_n\  \Omega^{a_1a_2} \cdots \Omega^{a_{2n-1}a_{2n}}
\epsilon_{a_1 \dots a_{2n}}
\end{equation*}
This quantity is equally mute about local information. By analogy,
in $d>2n$, the quantity $f(\Omega^{\wedge n}\wedge E^{d-2n})$ is
known as the dimensionally continued Euler
density~$\cite{Zumino-86,Teitelboim,BTZ}$. When written in terms of
tensors, the dimensionally continued density takes the same form as
the Euler density,
\begin{gather}\label{Euler density_tensor}
 \frac{1}{2^n}\ \delta^{\mu_1 \dots \mu_{2n}}_{\nu_1 \dots \nu_{2n}}
 R^{\nu_{1}\nu_2}_{\quad \mu_1 \mu_2} \cdots R^{\nu_{2n-1}\nu_{2n}}_{\qquad
  \mu_{2n-1}\mu_{2n}} \sqrt{-g}
\end{gather}
except that the dummy indices run over more values. In the vielbein
notation, the difference is more clear- the vielbeins appear
explicitly in the dimensionally continued density. The Lagrangian
formulation of Lovelock gravity involves a sum of terms which are
dimensionally continued Euler densities and yields the Lovelock
equations of motion~$\cite{Lovelock}$. This is a useful way to think
of the Lagrangian. This similarity to the Euler density accounts for
the interesting properties of the theory mentioned in the previous
paragraph.

In this paper, we deal with singular sources of gravity, that is
matter whose internal structure is restricted in dimensionality
lower than of that of the manifold. It is known that, of all
singular sources in Einstein's theory, the codimension 1
source~\cite{Israel,Geroch} is especially easy to describe
mathematically. The stress-energy-momentum tensor is unambiguously
well defined as a distribution. It has recently been realised that
this is also true for codimension 1 hypersurfaces in Lovelock
theory~\cite{Deruelle}. It is also known that, due to the
nonlinearity of Einstein's theory, there are problems and
ambiguities in describing singular sources of codimension greater
than $1$~\cite{Geroch}, although there is some hope of being able
to describe codimension $2$ sources in a meaningful
way~\cite{Garfinkle,Fursaev-95}. Just as point charges are useful
in studying electromagnetism, it is also useful to have a well
defined description of singular sources of gravity. Even if
singular sources do not exist as fundamental particles, they can
be useful as simple approximations.

Hypersurfaces of codimension one (hereafter just called
hypersurfaces), will generally intersect each other. It is then a
natural step to consider intersections. In ref.~\cite{Gravanis} we
found that there could be a singular energy-momentum tensor located
at intersections without any mathematical problems or ambiguities
(in particular, the vielbein frame is well defined at the
intersections). In the order $n$ Lovelock gravity, the singular
matter can live on intersections of codimension $n$ or less. Some
examples of intersections in Lovelock gravity have also been given
in refs. \cite{Navarro}.

Before going into the details of intersections, let us first
consider a hypersurface and the junction conditions~\cite{Israel}.
At a junction, the metric is continuous but the normal derivative
jumps. The part of the curvature that is intrinsic to the junction
is single valued but the extrinsic curvature representing the
embedding of the surface into the manifold is different on each
side. In the vielbein language it is the connection one-form that
is discontinuous. The problem then is that there are discontinuous
forms meeting at intersections. We would like to re-express the
problem in terms of continuous connection 1-forms so as to use
usual methods of differential geometry. To give a specific
example, consider the so called Einstein-Gauss-Bonnet theory in
five dimensions. There are three terms: the cosmological constant,
Einstein-Hilbert and a curvature squared term:
\begin{gather*}
 {\cal L} = \alpha_0 E^a \wedge E^b \wedge E^c \wedge E^d \wedge E^e
 \epsilon_{abcde}
 + \alpha_1 \Omega^{ab}\wedge E^c \wedge E^d \wedge E^e
 \epsilon_{abcde}
 + \alpha_2 \Omega^{ab}\wedge\Omega^{cd} \wedge E^e
 \epsilon_{abcde}
\end{gather*}
 Suppose that we have a single hypersurface $\Sigma$
which divides our space-time into two regions $M_-$ and $M_+$, such
that the metric is continuous but the connection may be
discontinuous. The junction conditions can be obtained by including
the surface term in the action:
\begin{gather}\label{EGB_surface}
 \alpha_1\int_{\Sigma}  (\omega_+^{ab} - \omega_-^{ab})\wedge E^c
 \wedge E^d \wedge E^e
 \epsilon_{abcde}\\\nonumber
 + \ \alpha_2\int_{\Sigma}   (\omega_+^{ab} - \omega_-^{ab})\wedge
 \left(\Omega_+^{cd} + \Omega_-^{cd} - \frac{1}{3} (\omega_+ -
 \omega_-)^c_{\ f}\wedge
 (\omega_+ - \omega_-)^{fd}\right)\wedge E^e
 \epsilon_{abcde}
\end{gather}
The Euler-Lagrange variation with respect to the connection
cancels the total derivative term coming from the bulk. The Euler
variation with respect to the vielbein gives a tensor which, when
set equal to the intrinsic stress tensor on $\Sigma$, gives the
correct junction conditions~\cite{Davis,Gravanis-02}. It is usual
to introduce the intrinsic connection $\omega_\|$, and the
corresponding curvature $\Omega_\|$ and second fundamental form
$\theta^{ab} := \omega^{ab}- \omega_\|^{ab}$. Then the surface
term would have the form:
\begin{gather*}
 \int_{\Sigma}  \alpha_1 \left[\theta^{ab}\wedge E^c
 \wedge E^d \wedge E^e
 \epsilon_{abcde}\right]^+_- + 2\alpha_2 \left[\theta^{ab}\wedge \left(\Omega_\|^{cd}
 + \frac{1}{3}\theta^{c}_{\ f}\wedge\theta^{fd}\right)\wedge E^e
 \epsilon_{abcde}\right]^+_-
\end{gather*}
where the brackets signify the jump in this surface term across the
boundary: $[f(\theta)]^+_- := f(\theta_+) -f(\theta_-)$. The first
term is the jump in the Gibbons-Hawking boundary term (first
written, it seems, by York~\cite{York}), written in terms of
differential forms. The second term is the jump in the boundary term
of Myers~\cite{Myers}. It is of course natural that the boundary
term which makes the action well defined on a manifold with boundary
also plays a role in the junction conditions.

The advantage of introducing the intrinsic connection into the
action is that one can write everything in terms of extrinsic and
intrinsic curvature tensors on $\Sigma$. One can then evaluate the
junction conditions using bulk metrics written in different
coordinate systems on each side. The intrinsic curvature is then
written in terms of the intrinsic reference frame on $\Sigma$. Then
the difference is taken between the two terms. In this way one can
avoid the dangers of confusing a real discontinuity with a purely
co-ordinate effect~\cite{Israel}. This approach is important if one
is calculating in terms of tensors, but not so essential if one uses
differential forms, a point which will be discussed in more detail
at the end of section \ref{Continued_section}. In the following we
shall introduce surface terms which, like (\ref{EGB_surface}), do
not explicitly contain the induced connections on the hypersurfaces.
In section \ref{Junction_section} we will return to make contact
with the formulation in terms of extrinsic curvature.

The question arises, can one describe intersections of hypersurfaces
in the same style, with boundary terms?
 The key point in ref.~\cite{Gravanis} is
that this can be done, thanks to the relationship of each term in
the Lagrangian to its topological cousin, the Euler density.

Now the Euler number is something that is actually independent of
the local form of the metric and associated connection,
\begin{gather}
\int_{M^{(2n)}} f(\Omega^{\wedge n}) = \int_{M^{(2n)}}
f((\Omega')^{\wedge n})\ \propto \text{Euler no}.
\end{gather}
It is a purely topological number. If we have a whole family of
(metric respecting) connections over $M$, $\omega_i$, related to
each other by homotopy, one can equally well write the Euler
number in terms of any of them. Also, and the important point for
us, one can formally rewrite the Euler number in terms of a
discontinuous connection, which coincides with each $\omega_i$ in
some region $i$ of $M$. One will then have boundary and
intersection terms in the integral. This amounts to a cellular
decomposition of the manifold into a honeycomb-type lattice. Note,
we do not localise curvature at the intersections (which would
introduce the complication of the connection picking up a gauge
transformation going round the singularity).

 The set of boundary and intersection terms were found
in  the previous work and are summarised in section 2. We introduced
a connection which interpolated between each $\omega_i$ by means of
some variables usually denoted by $t$ which we shall call homotopy
parameters. We found that each intersection term was a density built
from the curvature of the interpolating connection, integrated over
the homotopy parameters. In section 3, we shall re-derive these
results by a more geometrical method. We introduce a manifold, $W$,
which is locally a Cartesian product of each intersection and a
simplex in the homotopy parameters. We then introduce a closed form
$\eta$ in the space $W$. The closure of $\eta$ implies our
composition rule (\ref{comprule}), in other words the closure of
$\eta$ is sufficient to provide the composition rule. The results
can be presented in a simpler way by introducing a multi-parameter
generalisation of the Cartan homotopy operator. We should note that
these results are essentially the same as results found by Gabrielov
et al. in seeking a combinatorial formula for Characteristic
Classes~\cite{Gabrielov}.

The entire honeycomb formed out of the complicatedly intersecting
hypersurfaces is described by a few simple equations. All sorts of
intersections which it contains are accommodated in the scheme given
by these equations and the shape of $W$. For the Euler density, we
find an explicit expression for $\eta$ and show that it is closed.
The form of the intersection terms is clarified greatly.

In section 4 we turn to the action built from the dimensionally
continued Euler densities, where the vielbein enters explicitly into
the action. If there are hypersurface sources, the connection
$1$-form is discontinuous. Can we still rewrite the action in terms
of the continuous connection in each bulk region plus boundary
terms? We will show that the answer is yes and that the
gravitational intersection Lagrangians obey the same composition
rule (\ref{comprule}). This is because the dimensionally continued
$\eta$ is still closed on $W$.

In ref.~\cite{Gravanis}, we  generalised the intersection terms to
the dimensionally continued Euler densities in a natural way. The
resulting action was found to be one-and-a-half order in the
connection: the field equations come from independent variation of
vielbein and connection. The zero torsion constraint makes the
field equation from the variation of the connection vanish
identically. We thus concluded that this was the correct action,
the explicit variation with respect to the vielbein giving the
junction conditions for intersecting hypersurfaces in Lovelock
gravity. The key results of section $5$, Propositions $(6)$ to
$(9)$, verify this.

We can write the action which generates all the intersection terms
as
\begin{gather}
S = \int_W \eta
\end{gather}
$\eta$ is given by (\ref{final?}) for the Euler density and
(\ref{etadc}) for the dimensionally continued Euler density. The
validity of the formula in both the topological and the
gravitational case is the main result of this paper.

 In section \ref{Junction_section} the junction conditions for
intersections and collisions are elaborated in more detail and some
physical applications are discussed.

\section{The composition rule}

We will review the argument of ref.~\cite{Gravanis}. Let $\omega$ be
any connection and $\Omega$ the curvature. The continuous variation
of an invariant polynomial
\begin{gather}\label{POmega}
P(\Omega)= f(\Omega^{\wedge n})
\end{gather}
with respect to the connection produces the well known formula
\begin{gather}\label{basic transgression}
P(\Omega)-P({\Omega'})=d\, T\!P(\omega,\omega')
\end{gather}
where $T\!P$ is the Transgression of $P$~\cite{Chern}. This was
generalised to the composition formula:
\begin{equation}\label{comprule}
\sum_{s=1}^{p} (-1)^{s-p-1} {\cal L}(\omega_0,..,
\widehat{\omega_s},..,\omega_{p}) =d{\cal L}(\omega_0,..,\omega_{p})
\end{equation}
${\cal L}(\omega) = P(\Omega)$, ${\cal L}(\omega_1, \omega_2) =
T\!P(\omega_1,\omega_2)$ and an expression for the general ${\cal
L}$ was found. It was shown that these forms live on the
intersections of regions of $M$. Let's divide $M$ up into a
honeycomb of regions labelled by $i$ and denoted by $\{i\}$, with
intersections denoted by $\{ij\}$, $\{ijk\}$ etc, which are symbols
fully anti-symmetric in their indices and keep track of the
orientation. The integral over the manifold of ${\cal L}(\omega)$
can be rewritten
\begin{gather}\label{rewrite}
\int_M {\cal L}(\omega) = \sum_i \int_{\{i\}} {\cal L}(\omega_i)
+\sum_{k \geq 2} \frac{1}{k!}\sum_{i_1...i_k}\int_{\{i_1...i_k\}}
{\cal L}(\omega_{i_1},...,\omega_{i_k})
\end{gather}
Explicit formulae for these intersection terms were found. Each
intersection contributes to the action a term:
\begin{gather}\label{d-dimint}
\int d^{d-p}x \int d^pt \: \:  funct. \left(\omega(t)\right).
\end{gather}
where $t$ are the homotopy parameters and $\omega(t)$ interpolates
between the $\omega_i$'s. We will find somewhat simpler expressions
for these terms in the next section.

This composition rule applies to any invariant polynomial, such as
the Pontryagin Class. Because of our interest in Lovelock gravity,
we shall only discuss here the Euler density. The connection
$\omega$ is always the Lorentzian (or Riemannian) connection and
torsion free.

In the dimensionally continued case the left hand side of
(\ref{rewrite}) with the ${\cal L}(\omega_{i_1},...,\omega_{i_k})$
being replaced by the their dimensionally continued analogues, is
the gravitational action describing the system involving
intersecting hypersurfaces. The variation with respect to the
connections gives identically zero under zero torsion and the
variation with respect to the vielbein provides the junction
conditions. The existence of these intersection Lagrangians, coming
from the discontinuities of the connection, shows that, in order to
have a general treatment, one should allow for the possibility of
distributional parts of the matter's energy tensor with support at
the discontinuities
\begin{equation}\label{energy tensor with delta functions}
T=\sum_i  T_i \: f_i +\sum_{k \geq 2} \frac{1}{k!}\sum_{i_1...i_k}
T_{i_1...i_k} \: \delta(\Sigma_{i_1 \dots i_k})
\end{equation}
where  $T_i$ is the energy tensor of the region labelled by $i$ and
$f_i$ is a function that taken on the value 1 in the respective
region and 0 elsewhere; $T_{i_1 \dots i_k}$ is the energy tensor of
the intersection hypersurface $\Sigma_{i_1 \dots i_k}$, the
point-set whose orientation as embedded in $\Sigma_{i_2 \dots i_k}$,
$\Sigma_{i_1i_3 \dots i_k}$, etc, has been taken into account by the
fully anti-symmetric symbol $\{i_1 \dots i_k\}$. $\delta(\Sigma_{i_1
\dots i_k})$ is the delta function with support $\Sigma_{i_1 \dots
i_k}$.

Note that lower case Latin indices from the middle of the alphabet
label the bulk regions, not to be confused with space-time indices.

\section{A geometrical approach}\label{Geomsection}

This and the next section will be devoted to analyzing the purely
topological case, since part of our arguments can be understood in
this context. It will then help us to see what kind of refinements
are needed to pass to the gravitational case.

\begin{figure}
\begin{center}\mbox{\epsfig{file=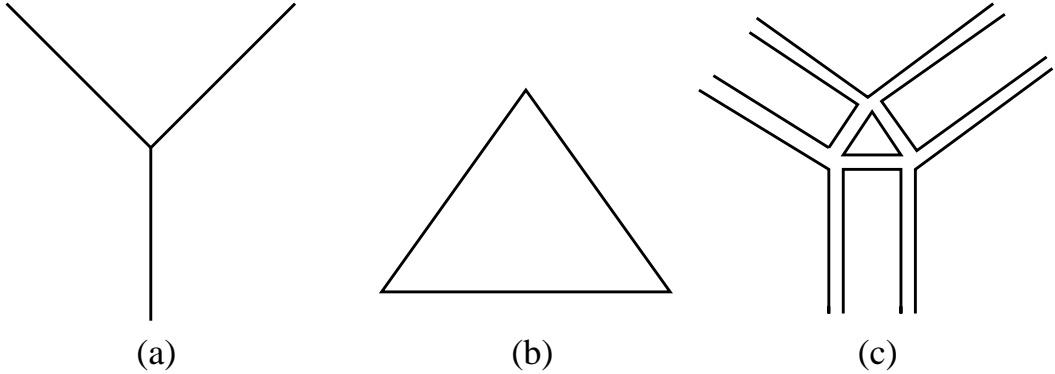, angle=-90, width=14cm}}
\caption{{\small (a) Simplicial intersection; (b) Simplex in the
space of homotopy parameters; (c) A projected diagram of the space
$W$ (equation (\ref{Walanerve})). Every $d-1$ dimensional surface is
`thickened' in the $t$-space by a 1-dimensional simplex. These meet
at a $d-2$ dimensional intersection, which is ``thickened" by a
triangle in the $t$-space.}}\label{simp}
\end{center}
\end{figure}

We want to describe the situation in the vicinity of an intersection
of codimension $p$ between different bulk regions. In this vicinity
there will also be intersections of lower codimension. At each
intersection, we have a meeting of connections $\omega_i$ in the
different regions. Let us for the moment deal only with simplicial
intersections: we define the simplicial intersection of codimension
$p$ to be a surface of codimension $p$ where $p+1$ regions meet
(fig. \ref{simp}(a)). It was found in ref.~\cite{Gravanis} that the
${\cal L}(\omega_0,...,\omega_{p})$ is an integral over $p$
different homotopy parameters interpolating between the connections
(see (\ref{d-dimint})).

If we look at (\ref{d-dimint}) we make the following observation:
each order of intersection causes us to lose a dimension but gain an
extra connection. Each new connection means an extra parameter of
continuous variation.
With this in mind, we can think of our action
as an integral over a $d$-dimensional space which is a mixture of
space-time and $t$ directions.

Let us interpolate in the most symmetrical way. We introduce a
$N$-dimensional simplex in the Euclidean space $\mathbf{R}^{N+1}$
with coordinates $t$. Let us define the interpolating connection:
\begin{gather}
\omega(t) := \sum_{i=0}^{N} t^i\ \omega_i, \qquad \sum_{i=0}^{N} t^i
=1
\end{gather}
and the associated curvature:
\begin{gather}
\Omega(t) := d \omega(t) + \frac{1}{2}[\omega(t),\omega(t)].
\end{gather}
So we introduce the space $F =  S_{N} \times M$, with $S_{N}$, a
simplex of dimension $N$. The latter will be frequently called
$t$-space. Each of $N$ points of the simplex corresponds to a
continuous connection form $\omega_i$ on M with its support on some
open set in $M$ containing the region $i$; we add another one, that
is, one more connection for reasons that will become clear later.
Each contribution to our action will live on some $d$-dimensional
subspace of $F$. The technical reason for introducing the space $F$
is that the connection is continuous on it and integration is well
defined. There is also an aesthetic reason. It is quite a nice
feature of the problem that the mathematics will take on its
simplest form when the $t$-space is a simplex, as we have already
chosen. It provides a geometrical picture which can simplify many
calculations. For example, the treatment of a non-simplicial
intersection becomes easy, as we shall see in below.

Let us define a $d$-dimensional differential form in this space $F$
(where for convenience the $dx$'s are suppressed).
\begin{gather}\label{zetaform}
\eta := \sum_{l=0}^{n} \frac{1}{l!}\,
dt^{i_1}\wedge\cdot\cdot\cdot\wedge dt^{i_l}
\wedge\eta_{i_1...i_l}(x,t),
\\\nonumber\eta_{i_1...i_l}:=\eta_{i_1...i_l
\mu_{l+1}...\mu_{d}}dx^{\mu_{l+1}}\wedge\cdot\cdot\cdot\wedge
dx^{\mu_d}.
\end{gather}
We can now proceed to integrate this form over different faces of
$S_{N}$. A $p$-face (which we call $s_{0...p}$ or just $s$) is a
subsimplex of $S_{N}$ which interpolates between a total of $p+1$
different connections(fig. \ref{simp}(b) ):
\begin{gather}
s_{0\dots p} = \Big\{\Big.(t^0,\dots,t^p,0,\dots,0)\,\Big|\, t^i
\geq 0 ,\, \sum_{i=0}^p t^i =1\Big\}.
\end{gather}
Let us define ${\cal L}_{0...p}$  to be the integral over the
$p$-dimensional face:\footnote{Strictly there should be a factor of
$(-1)^{P(0,...,p)}$ in the middle term to account for the
orientation with respect to $S_{N}$. However, we can choose $s$ to
have the positive orientation by assuming the points $0...p$ are in
the appropriate order.}
\begin{gather}\label{defLeta}
{\cal L}_{0...p} := \int_{s_{0...p}} \eta =
\frac{1}{p!}\int_{s_{0...p}} dt^{i_1}\cdot\cdot\cdot dt^{i_{p}}
\eta_{i_1...i_{p}},
\end{gather}
$\eta$ here being understood to be the restriction of $\eta$ onto
$s$ so that the integral is a function of $x$ only. This integral
picks out terms in $\eta$ which are a volume element on the
appropriate face. We would like, for an appropriate choice of
$\eta$, to identify this term with ${\cal L}(\omega_0,...,\omega_p)$
as defined in the introduction with respect to the Euler density. We
shall see that this can indeed be done and we shall find a simple
form for $\eta$.
\\\\
{\bf Proposition (1)}: A sufficient condition on $\eta$ such that
${\cal L}_{0...p}$ obeys the composition rule (\ref{comprule}) is
that $\eta$ be a closed form, $d_F \eta = 0$. Here the exterior
derivative on F is $d_F = d_{(x)} + d_{(t)}$. $d_{(t)}$ and
$d_{(x)}$ are the exterior derivative restricted to the simplex and
to $M$ respectively.
\\\\
{\bf Proposition (2)}: The form of $\eta$ corresponding to the Euler
density is
\begin{gather}\label{final?}
\eta = f\Big(\big[d_{(t)}\omega(t)+\Omega(t)\big]^{\wedge n}\Big).
\end{gather}
This formula is already known in the mathematics
literature~\cite{Gabrielov}. $\eta$ is closed on $F$: $d_F
\eta=0$.
\\\\
{\bf Proposition (3)}: The intersection Lagrangian can be recovered
by the specific choice:
\begin{gather}\label{expliciteta}
\eta_{1...p} = A_p
f\big(\chi_1\wedge...\wedge\chi_p\wedge\Omega(t)^{\wedge(n-p)}\big),\\
\nonumber A_p =(-1)^{p(p-1)/2}\frac{n!}{(n-p)!}\\\label{Lagrange}
\Rightarrow {\cal L}(\omega_0,...,\omega_{p}) = A_p \int_{s_{01..p}}
d^{p}tf\big(\chi_1\wedge...\wedge\chi_p
\wedge\Omega(t)^{\wedge(n-p)}\big).
\end{gather}
$\chi_i \equiv \omega_i - \omega_0$ and $\omega(t) = \omega_0 +
\sum_{i=1}^p t^i\chi_i$. The intersection Lagrangian is zero for
$p>n$.
\\

To prove the first proposition, we will need to use Stokes' Theorem
on the face $s$.
\begin{gather}\label{tinteg}
\int_s d_{(t)} \eta = \int_{\partial s} \eta.
\end{gather}
The boundary of the simplex $s_{0...p}$ is
\begin{gather}
\partial s_{0...p} = \sum_{i=0}^p (-1)^{i} s_{0...\widehat{i}...p}
\end{gather}
with the orientation being understood from the order of the indices.

Now let us integrate the form $d_{(x)}\eta$ over the face. We will
need to remember that in permuting this exterior derivative past the
dt's we will pick up a $\pm$ factor.
\begin{gather}
d_{(x)}\eta = \sum_l\frac{(-1)^l}{l!}\,dt^{i_1}
\wedge\cdot\cdot\cdot\wedge dt^{i_l} \wedge
d_{(x)}\!\eta_{i_1...i_l}.
\end{gather}
Using this information we may integrate over the $p$-face $s$
\begin{gather}\label{xinteg}
\int_s d_{(x)}\eta = (-1)^{p} d\int_s \eta.
\end{gather}

Combining equations (\ref{tinteg}) and (\ref{xinteg})
\begin{gather}\label{intdFeta}
\int_{s_{0...p}} d_F \eta = (-1)^{p} d\int_{s_{0...p}} \eta +
\sum_{i=0}^p (-1)^{i} \int_{s_{0...\widehat{i}...p}} \eta.
\end{gather}
If our form $\eta$ is closed in F, $d_{F}\eta$ must necessarily
vanish term by term in the $dt$'s and $dx$'s. The integral on the
right hand side of (\ref{intdFeta}) must therefore vanish. Recalling
the definition (\ref{defLeta}) we have proved Proposition 1:
\begin{gather}\label{equiv}
d_{F}\eta =0\quad \Rightarrow\quad d{\cal L}_{0...p}=\sum_{i=0}^{p}
(-1)^{p-i-1} {\cal L}_{0...\widehat{{i}} ...p}\ .
\end{gather}
The condition that $\eta$ be closed indeed implies our composition
formula.

The proof of Proposition 2 and 3 is in the appendix. We now turn to
discuss the space where the form $\eta$ lives and write down the
topological invariant in the presence of connection discontinuities.

\section{$W$ space and the action of the system}

We have seen that the simplicial intersection is related to a
simplex in the parameter space. The simplex with $p+1$ vertices
corresponds to a codimension $p$ intersection with $p+1$ bulk
regions meeting. The example $p=2$ is shown in fig. 1(a)- the
intersection looks like the simplex turned inside out. As pointed
out already, the connection $\omega(t)$ is smooth on $F$ where the
$d$-dimensional Lagrangian density $\eta$ is defined.

Now define a $d$-dimensional space $W \subset F$ as follows.
Consider a simplicial intersection and set
\begin{equation}\label{defW}
W=\sum_{p=0}^{h}\sum_{i_0\ldots i_p} \frac{1}{(p+1)!}\, s_{i_0\ldots
i_p} \times \{{i_0\ldots i_p}\} \: \subset F
\end{equation}
where $h$ is the highest codimension of the intersections present.
Let's write the first few terms explicitly
\begin{equation}
W=\sum_i s_i \times \{i\}+\sum_{i<j} s_{ij} \times
\{ij\}+\sum_{i<j<k} s_{ijk} \times \{ijk\}+ \dots
\end{equation}
The first term contains the bulk regions multiplied with
0-simplices, the second contains the hypersurfaces multiplied with
1-simplices, the third contains the codimension 2 intersections
multiplied with the associated 2-simplex, and so on. All the
intersection sub-manifolds of $M$ have been put in a single
$d$-dimensional space, where again $d={\rm dim}M$. Moreover this is
a $d$-dimensional space where the connection $\omega(t)=\sum_i t^i
\omega_i$ is \emph{continuous}.


Define the curvature associated with the connection $\omega(t)$ over
$W \subset F$, that is, the derivative operator is $d_F$ :
\begin{equation} \label{defomegaF}
\Omega_F := d_F \omega(t)+ \frac{1}{2}[\omega(t),\omega(t)]
\end{equation}
From that we have
\begin{equation}
\Omega_F=d_{(t)} \omega(t) +\Omega(t)
\end{equation}
Being a curvature form it satisfies the Bianchi identity. To see
this explicitly let $D_F$ be the covariant derivative associated
with the derivative operator $d_F$ and connection $\omega(t)$. We
have
\begin{equation}\label{Fbianchi}
D_F\Omega_F=(d_{(t)}+D(t))(d_{(t)} \omega(t)
+\Omega(t))=d_{(t)}\Omega(t)+D(t)d_{(t)}\omega(t)+D(t)\Omega(t)=0
\end{equation}
as the first two terms cancel each other by
$d_{(t)}d_{(x)}=-d_{(x)}d_{(t)}$ and the last by the Bianchi
identity of $\Omega(t)$. This is discussed in more detail in the
appendix A.

 The action takes the simple form on $W$ :

{\bf  Proposition (4)}: The action of the whole system takes the
form
\begin{gather}\label{etaisEuler} S= \int_W \eta, \qquad \eta =
P(\Omega_F)
\end{gather}
\emph{Proof} : It is an immediate consequence the form of action
(\ref{rewrite}) derived in~\cite{Gravanis} and of the definition
of $W$ above under the consistent identification (\ref{defLeta})
according to Propositions 2 and 3.\footnote{It is an abuse of
terminology to speak about `action' in the purely topological
case. In the next section it will become clear why the real
gravitational action is found in almost the same way.} We return
to prove this explicitly in the next section, for the more general
case of the dimensionally continued density by the methods
introduced in this paper.

 The manifold $W$ is further
discussed in appendix \ref{topappendix} where some discussion of
topology is also included. There it is proved the following

 {\bf Proposition (5)}: If $\partial M=0$ then
\begin{equation}
\partial_F W=0.
\end{equation}
In fact it is shown as being equivalent to the definition of the
simplicial intersection's boundary rule introduced
in~\cite{Gravanis}. This is a homological version of the
statements of the previous section. Also this proposition helps us
prove easily the invariance of the quantity $\int_W P(\Omega_F)$
under continuous variations of the connection: under $\omega(t)\to
\omega(t)+\delta\omega$ we have easily have
$\delta\Omega_F=D_F\delta\omega$ so
\begin{equation}\label{topo}
\delta\int_W P(\Omega_F)=\int_W n f(D_F\delta\omega\wedge
\Omega_F^{\wedge(n-1)})=\int_W
nd_Ff(\delta\omega\wedge\Omega_F^{\wedge(n-1)})=0
\end{equation}
where in the second equality we used the invariance of $f$ (see
appendix A) and and the Bianchi identity (\ref{Fbianchi}) and in the
last step the previous Proposition. In the next section this will
translate to a well-defined variational principle of a gravity
action i.e. a functional providing equations of motion for $E$ and
$\omega$.

The proof of Proposition 1 can be thought of in terms of a
generalisation of Cartan homotopy formula to a higher number of
homotopy parameters. Let the operator $K_s$ be defined by $K_s \eta
:= \int_s \eta$ and let $K_{\partial s} := \int_{\partial s}\eta$.
The equation (\ref{intdFeta}) can be written as
\begin{gather}\label{compisCartan}
(K_s \cdot d_F-(-1)^m d_F\cdot K_s)\eta = K_{\partial s}\eta
\end{gather}
This reduces to the usual Cartan homotopy formula for the 1-simplex
$m=1$.
\begin{gather}
(K_{01} d_F+d_F K_{01})\eta = \eta(1) -\eta(0)
\end{gather}
In fact, the whole of our analysis can be reduced down to the two
equations (\ref{etaisEuler}) and (\ref{compisCartan}). In words :
the whole intersection Lagrangian is a density in some manifold
which is locally a product of space-time and a simplex. The
composition rules are an expression of this higher dimensional
Cartan homotopy operator acting on this density.

Now the non-simplicial intersection in $M$ (that is, when $k
> p$ regions meet at a codimension $p$ surface) can be treated quite
easily. Instead of integrating over a simplex one integrates over a
simplicial complex in $t$-space. More than one face of dimension $p$
in $S_{N}$ are associated with the same $(d-p)$-surface in $M$.

\begin{figure}
\begin{center}\mbox{\epsfig{file=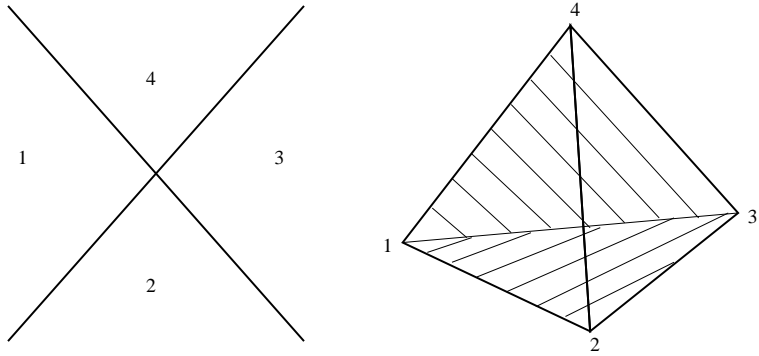, width=10cm}}
\caption{Non-simplicial Intersection. (a) Space-time,
(b)t-space.}\label{non-simp}
\end{center}
\end{figure}

Let us consider a simple example. We have four regions, 1,..,4,
meeting at a codimension 2 intersection $I\subset M$ which is has no
boundary (fig.\ref{non-simp}).
There are four hypersurfaces $\{12\},\{23\},\{34\}$ and $\{41\}$
meeting at $I$. Now look for a 2-chain $c$ such that
\begin{equation}\label{bchain}
\partial c=s_{12}+s_{23}+s_{34}+s_{41}
\end{equation}
as the r.h.s. is a 1-cycle, that is a 1-chain annihilated by the
boundary operator $\partial$. One solution to this equation is
\begin{equation}
c=s_{123}+s_{341}
\end{equation}
which is clearly not unique: (\ref{bchain}) tells us that a new
chain differing from the old one with a boundary is another
solution. If for example we choose a new 2-chain $c'$
\begin{equation}
c'=c+\partial  s_{1234}=s_{234}+s_{124}
\end{equation}
still $\partial c'=s_{12}+s_{23}+s_{34}+s_{41}$.

Integrating $d_F \eta=0$ over the two sides of (\ref{bchain}) we
have
\begin{align}
\int_{s_{12}+ s_{23}+ s_{34}+ s_{41}}\eta =&d\int_{c}\eta \nonumber\\
\Rightarrow {\cal L}_{12}+{\cal L}_{23}+{\cal L}_{34}+{\cal L}_{41}
=& d\left({\cal L}_{123}+{\cal L}_{341}\right)
\end{align}
According to what we have seen in~\cite{Gravanis} building a
functional with a well defined variational principle in the
presence of intersections, the last equation tells us that the
appropriate term for the non-simplicial intersection $I \subset M$
is
\begin{gather}\label{Ilagr}
\int_{I}\int_{c} \eta=\int_I \left({\cal L}_{123}+{\cal
L}_{341}\right)
\end{gather}
This is a special case of the result obtained in \cite{Gravanis} by
more conventional means, and can of course be easily checked in our
new language: the $W$ space reads
\begin{equation}
W=\sum_{i=1}^4 s_i \times
\{i\}+s_{12}\times\{12\}+s_{23}\times\{23\}+s_{34}\times\{34\}+s_{41}\times\{41\}+c\times
I
\end{equation}
with
$\partial\{21\}=\partial\{32\}=\partial\{43\}=\partial\{14\}=-I$ and
$\partial I=0$. Then
\begin{equation}
\partial_F W=\left(-(s_{12}+ s_{23}+ s_{34}+ s_{41}) +\partial c \right) \times I
\end{equation}
which vanishes for any chain $c$ satisfying (\ref{bchain}). So,
recalling (\ref{topo}), $\int_W \eta$ has a well-defined variational
principle. The action term at $I$ should be given by $\int_I\int_c
\eta$.

This action term is unique: the arbitrariness $c \to c+\partial c'$,
for any 3-chain $c'$, does not affect the action as the term above
changes by
\begin{equation}
\int_I \int_{\partial c'} \eta=\int_I \int_{c'} d_{(t)} \eta=-\int_I
\int_{c'} d_{(x)} \eta=\int_{\partial I}\int_{c'} \eta=0
\end{equation}
where we used
  $0=d_F
\eta=d_{(t)}\eta+d_{(x)}\eta$ in the second equality and $\partial
I=0$ in the last equality and the usual rule of commutation of a
$t$-space integral with $d_{(x)}$ operator.

\section{Dimensionally continued Euler
density}\label{Continued_section}

So far we have been considering the topological density. This is not
much good as a Lagrangian. We know that the action yields no
equations of motion. The point is that we can apply what we have
learned to the dimensionally continued densities. The Lovelock
Lagrangian (\ref{Shovelock}) is a combination of such densities. Now
we assume that the connection is a metric compatible (Lorentz)
connection. There are now explicit factors of the vielbein frame
$E^a$ appearing in the action.

We have a manifold $M$, of dimension $d$, with regions labelled by
$i$, separated by surfaces of matter. The vielbein $E$ is continuous
but the connection $1$-form $\omega$ is discontinuous at the
surfaces. Once again we rewrite the Lagrangian in terms of the
continuous connections $\omega_i$ and boundary terms. We interpolate
as before:
\begin{gather}
E(t) := \sum_{i=0}^{N-1}t^i\ E_i +t^N E,\qquad \omega(t) :=
\sum_{i=0}^{N-1} t^i\ \omega_i+t^N \omega.
\end{gather}
and again we define the space $F=S_{N}\times M$. As well as the
$d$-dimensional manifold $W$, defined in (\ref{defW}), we also
introduce the $(d+1)$-dimensional manifold:
\begin{gather}
W^+ = \sum_{p=0}^{h}\sum_{i_0\ldots i_p} \frac{1}{(p+1)!}\,
s^+_{i_0\ldots i_p} \times \{{i_0\ldots i_p}\} \subset F,
\\
s^+_{i_0\ldots i_p} =
\Big\{\Big.(t^0,\dots,t^p,0,\dots,0,t^N)\,\Big|\, t^i \geq 0 ,\,
\sum_{i=0}^p t^i+t^N =1\Big\}.
\end{gather}
The difference from the previous section is that we include the
physical vielbein and connection, $E$, $\omega$, as well as the
$E_i$, $\omega_i$.

We impose the two constraints:

\noindent  \verb"i") The vielbein frame is continuous across $M$. At
an intersection $\{i_0\dots i_p\}$:
\begin{equation*}
 E_{i_0}|_{\{i_0\dots i_p\}} =
\dots = E_{i_p}|_{\{i_0\dots i_p\}} = E|_{\{i_0\dots i_p\}}
\end{equation*}
 \verb"ii") Each connection is torsion-free:
\begin{equation*} d_{(x)}E_i^a +{\omega_i}^a_{\ b} \wedge  E_i^b =0
\end{equation*}

In fact, a good alternative way to define $W$ is: $W$ is the region
in $F$ where $E(t,x) = E(x)$ (of course $d_{(x)}E(t,x)$ is a
function of $t$ because the derivative of the metric is
discontinuous on $M$). Let $\phi_+$ be the embedding $\phi_+: W^+
\to F$. Let $D(t)$ be the covariant derivative associated with
$\omega(t)$ and $d_{(x)}$. From the two constraints we derive:
\begin{align}
\phi_+^* (d_{(t)} E(t)) & = 0,\label{dE} \\
\phi_+^* D(t)E(t)^a & =  0.\label{DE}
\end{align}
To prove the second equation we have used constraints \verb"i" and
\verb"ii" as well as \begin{gather*} \sum_{i=0}^p t^i d_{(x)} E_i^a
+ \sum_i t^i {\omega_i}^a_{\ b}\wedge E(t)^b =\sum_i t^i
{\omega_i}^a_{\ b}\wedge (E(t)-E_i)^b.
\end{gather*}
By the definition of the covariant derivative $D_F=d_{(t)}+D(t)$,
(\ref{dE}) and (\ref{DE}) give
\begin{equation}
\phi_+^* (D_F E(t))  =  0.\label{DFE}
\end{equation}

By these the composition formula is unchanged. To see that this is
the case we make use of the invariance property of the polynomial
contracted with the epsilon tensor:
\begin{align}
\phi_+^*d_{F} f\Big((d_{(t)}\omega(t)+\Omega(t))^{\wedge(n)}\wedge
E(t)^{\wedge(d-2n)}\Big) & =\nonumber\\
\phi_+^*D_{F}f \Big((d_{(t)}\omega(t)+\Omega(t))^{\wedge(n)}\wedge
E(t)^{(d-2n)}\Big) & =  0.
\end{align}
This vanishes by (\ref{defomegaF}), (\ref{Fbianchi}) and
(\ref{DFE}).

So we can define the form, closed in $W^+$:
\begin{gather}\label{etadc}
\eta_{DC} = f\big((d_{(t)}\omega(t)+\Omega(t))^{\wedge(n)}\wedge
E(t)^{\wedge{(d-2n)}}\big)
\end{gather}
An alternative notation will be
\begin{equation}\label{}
\eta(\omega(t),E(t)) \equiv \eta_{DC}.
\end{equation}

 The intersection terms will be terms in the expansion of
$\eta_{DC}$ integrated over the appropriate simplex in $F$. We
define:
\begin{gather*}{\cal
L}(\omega_0,...,\omega_p,E_0,\dots,E_p) := \int_{s_{0\dots p}}
\eta_{DC}.
\end{gather*}

We can now state:

{\bf Proposition (6):} $\phi_+^* d_F \eta_{DC}=0$.\\

Further, by (\ref{intdFeta}), the composition rule for ${\cal
L}(\omega_0,\dots,\omega_p,E_0,\dots,E_p)$ applies, when restricted
a codimension $(p-1)$ intersection $\{1\dots p\}$, where $E_0 =
\cdots = E_p = E$.
\begin{multline}\label{funkycomprule}
\left.d{\cal L}(\omega,\omega_1,...,\omega_p,E)\right|_{\{1\dots
p\}}-\sum_{i=1}^{p} (-1)^{p-i-1} \left.{\cal
L}(\omega,\omega_1,\dots\widehat{\omega_{i}} \dots
\omega_p,E)\right|_{\{1\dots p\}} = \\  = (-1)^{p-1}\left.{\cal
L}(\omega_1,\dots, \omega_p,E)\right|_{\{1\dots p\}}.
\end{multline}
The connection $\omega$ is the physical (discontinuous) connection.
Each term on the left hand side is ill defined, but the sum of them
is formally equal to the right hand side.

For the dimensionally continued case, $\eta_{DC}$ and ${\cal L}$ are
no longer Euler densities. It was therefore not obvious that our
composition formula should survive. It does survive though because
$\eta_{DC}$ is still a closed form when restricted to  $W^+ \subset
F$.

As a consequence of the composition rule the infinitesimal variation
of the action
\begin{gather}
{\cal S} = \int_W \eta_{DC}
\end{gather}
with respect to the connection vanishes~\cite{Gravanis} (provided
we impose the torsion free condition on the connection and
continuity of the metric) and the equations of motion just come
from the explicit variation with respect to the vielbein.

We can prove this fact now in a neat way using the $W$ space. We
assume $M$ has no boundary. Then, according to appendix B,
$\partial_F W=0$. Under variation $\omega(t)\to \omega(t)+\delta
\omega$,
\begin{equation}\label{zeroomegavar}
\delta_{\omega} \int_W f(\Omega_F^{\wedge n} \wedge E(t)^{\wedge
(d-2n)})=n \int_W d_F f(\delta \omega \wedge \Omega_F^{\wedge(n-1)}
\wedge E(t)^{\wedge(d-2n)})=0.
\end{equation}
where we have used the invariance of $f$, the variation of
$\Omega_F$: $\delta \Omega_F=D_F \delta\omega$,  the identity
(\ref{Fbianchi}) and the constraint (\ref{DFE}) restricted on $W$.
So we have:

{\bf Proposition (7)}. Under continuity of the vielbein and the
torsion-free condition on each bulk connection (conditions \verb"i"
and \verb"ii" restricted on $W$) the field equations for the
connection are trivial: $\delta_{\omega} \int_W \eta_{DC}=0$.

We can also show that the various intersection terms in the action
do produce a diffeomorphism invariant action functional in the
presence of discontinuities.
Let 
 $\xi$ be
a vector field.
Then diffeomorphism invariance of the action $\int_W \eta_{DC}$ can
be expressed as
\begin{equation}
0=\delta_{\xi} \int_W \eta_{DC}=\int_W \pounds_{\xi} \eta_{DC}
\end{equation}
where $\pounds$ is the Lie derivative. But
\begin{equation}
\int_W \pounds_{\xi} \eta_{DC}=\int_W i(\xi) d_F \eta+d_F i(\xi)
\eta_{DC}
\end{equation}
where we express the Lie derivative via a well-known identity
involving the inner product operator $i(\xi)$ and used also
$i(\xi)d_{(t)}+d_{(t)}i(\xi)=0$ as $d_{(t)}\xi=0$. From $\partial_F
W=0$ the second term in the above equation vanishes and we have the
condition
\begin{equation}\label{simplediff}
\int_W i(\xi) d_F \eta_{DC}=0
\end{equation}
for all vector fields $\xi$. This implies that
\begin{equation}\label{}
\Big(\sum_{i=0}^p(-1)^{i-p}  {\cal
L}(\omega_0,\dots,\widehat{\omega_i},\dots,\omega_p,E)+ d{\cal
L}(\omega_0,\dots,\omega_p,E) \Big)\Big|_{\{01\dots p\}}=0
\end{equation}
at an arbitrary intersection chosen here to be $\{01\dots p\}$. This
is a quite different composition rule than (\ref{funkycomprule}).
Nevertheless it is true,
 as the quantity on the
l.h.s. of (\ref{simplediff}) does vanish by the conditions \verb"i"
and \verb"ii" (and the Bianchi identity (\ref{Fbianchi})).

Let us look again at the non-simplicial intersection. We have seen
that the arbitrariness in the choice of the chain $c$ in $\int_I
\int_c \eta$ does not affect the action in the purely topological
case because $d_F \eta=0$. The dimensionally continued density is
closed only when restricted to a subspace of $F$. In the specific
example we treated in the previous section, the arbitrariness $c \to
c+\partial c'$ corresponds to a change in $W$ space as $W \to
W+\partial_F Y$ where $Y=c' \times I$. That is, the action $\int_W
\eta_{DC}$ is unaffected if and only if
\begin{equation}\label{collapsibility}
\int_{\partial_F Y} \eta_{DC}=\int_Y d_F \eta_{DC}=0
\end{equation}
where $Y$ is a $(d+1)$-dimensional space. Let for example
$c'=s_{1234}$. The above equation is guaranteed by the fact that
continuity of the vielbein ensures that the pullbacks of
$d_{(t)}E(t)$ and $D(t)E(t)$ onto $Y$, with $E(t)=\sum_{i=1}^4 t^i
E_i$, vanish.

The composition rule (\ref{funkycomprule}) can be used to derive
that the action $\int_W \eta(\omega(t),E(t))$ is formally equivalent
to the action ${\int_M \cal L}(\omega,E)$ (Lemma 3 of
ref.~\cite{Gravanis}), provided there exists an everywhere
continuous vielbein frame $E$ and $DE = 0$.  We prove that now in a
more elegant and general way as follows.

{\bf Proposition (8).}  The relation
\begin{equation}\label{gravi-transgression formula}
\phi^* \eta(\omega,E)=\phi^* \eta(\omega(t),E(t))+\phi^* d_F {\cal
B}
\end{equation}
holds, for some differential form ${\cal B}$, provided that
conditions \verb"i" and \verb"ii" are satisfied. \\\\
\textit{Proof}: We interpolate by $\omega(t,s)=(1-s)\omega(t)+s
\omega$ and $E(t,s)=(1-s) E(t)+ s E $, with $0 \le s \le 1$. We can
easily show that $\phi^*(d_F E(t,s)^a+\omega(t,s)^a_{\ b}
E(t,s)^b)=0$. By the Chern-Weil procedure one finds that
(\ref{gravi-transgression formula}) holds and
\begin{equation}\label{}
{\cal B}=n \int_0^1 ds\:  f \big( (\omega-\omega(t)) \wedge
\Omega_F(s)^{\wedge (n-1)} \wedge E(t,s)^{\wedge(d-2n)} \big)
\end{equation}
where $\Omega_F(s):=d_F \omega(t,s)+\frac{1}{2}
[\omega(t,s),\omega(t,s)]$. We have then proved a gravitational
version of the transgression formula (\ref{basic transgression}). As
$\omega$ is not well defined at the hypersurfaces, formula
(\ref{gravi-transgression formula}) holds in the weak sense.
\\

We prove now the equivalence. We start by noting that:
\begin{gather}\label{obvious}
\int_M {\cal L}(\omega,E) = \int_W \eta (\omega,E).
\end{gather}
We integrate over $W$ the identity of Proposition (8). From the
Proposition (B1) we have that for $\partial M=0$ the space $W$ has
no boundary so
\begin{equation}\label{}
\int_W \eta(\omega,E)=\int_W \eta(\omega(t),E(t)).
\end{equation}
Therefore we have proved:

{\bf Proposition (9).} The action
\begin{gather}\label{final action in proposition}
{\cal S} =\int_W \eta(\omega(t),E(t))= \sum_i \int_i {\cal
L}(\omega_i,E) +\sum_{k \geq 2}
\frac{1}{k!}\sum_{i_1...i_k}\int_{\{i_1...i_k\}} {\cal
L}(\omega_{i_1},...,\omega_{i_k},E),\\\nonumber {\cal
L}(\omega_0,..,\omega_{p},E) = \sum_{n=p}^{[d/2]} \alpha_n A_p
\int_{s_{0...p}}\hspace{-.2in}
d^pt\,f\big((\omega_1-\omega_0)\wedge\!\cdots\wedge
(\omega_{p}-\omega_{0})\wedge \Omega(t)^{\wedge(n-p)}\wedge
E^{\wedge(d-2n)}\big),
\end{gather}
is formally equivalent to $\int_{M} {\cal L}(\omega,E)$.

 By Proposition (7) the equations of motion (junction conditions)
come only from variation with respect to the vielbein. Since the
action is algebraic in the vielbein they are now easily obtained.

\begin{center} *** \end{center}

It is worth noting the following. The action ${\cal S}$ and the
implied equations of motion involve explicitly only bulk data. One
can easily prove the formula
\begin{equation} \label{symOmegat} \Omega(t)=\sum_i t^{i} \Omega_{i}
-\frac{1}{4} \sum_{ij} t^i t^j
[\omega_i-\omega_j,\omega_i-\omega_j].
\end{equation}
One only needs to calculate the bulk connection jumps and bulk
curvature forms. The intrinsic connection on each hypersurface is
virtually absent from the formulas. It is only implicitly there by
continuity of the veilbein and the vanishing of torsion everywhere.
If there is a discontinuity at the codimension one submanifold
$\Sigma$, then the purely tangential part of the derived connections
is continuous and defines an intrinsic connection of $\Sigma$.
Consider a non-null $\Sigma$. Let an adapted frame $(E^N, E^{\mu})$
at $\Sigma$ where $E^N$ is normal and $E^{\mu}$ are tangential to
it. Then the pullback of the tangential torsion reads $i^*
dE^{\mu}+i^*\omega_{\ \nu}^{\mu} \wedge E^{\nu}=0$. If $E^{\mu}$ is
continuous across $\Sigma$ then $i^* \omega^{\mu}_{\ \nu}$ is a
natural intrinsic connection on it. The continuous components do not
actively participate in the interpolations $\omega(t)=\sum_{i=0}^n
t^i \omega_i$ i.e. it is absent in $d_{(t)} \omega(t)$ and it
cancels out in the differences $\omega_i-\omega_j$. Then, as we also
discuss in more detail below, these bulk connection differences are
simply the jump of the second fundamental form across the
hypersurfaces $ij$. In general one need not explicitly introduce the
Cauchy data of the hypersurface for the junction condition
calculation. The results are equivalent and the actual calculations
are often greatly facilitated.

Lack of need for intrinsic data in the junction conditions formulas
means that they are applicable to the null hypersurface as well.
This is also suggested by the fact that the inverse of the spacetime
metric $g$ appears nowhere. So if in particular the induced metric
becomes degenerate somewhere, as in the case of null hypersurfaces,
the formulas still hold.

Finally note that the vielbein may be more naturally given in a
different frame on each side of the hypersurface. In this case,
there is a set of vielbeins $E^{\mu}$ on one side and a different
set of vielbeins $E^{\hat{\mu}}$ on the other. If the induced metric
is the same then $E^{\mu}$ and $E^{\hat{\mu}}$ must be related by a
local Lorentz transformation across a hypersurface. Then the
connection also differs by a gauge transformation across the
hypersurface. In order to get the correct junction conditions, this
Lorentz transformation must be taken into account. In practice, it
may be more convenient to do calculations on each side of the
hypersurface with the respective natural intrinsic connection and
put together the results. The justification of the calculation is
based on the composition rule (\ref{comprule}) with intrinsic as
well as bulk connections involved, and it is an interesting problem
on its own to be discussed elsewhere. In any case the result can be
obtained from (\ref{final action in proposition}) by replacing $E$
and the respective $\omega$ with the transformed ones.

\section{Explicit junction conditions for
intersections}\label{Junction_section}

Let us make contact with more standard formulations of junction
conditions. We take the chance to comment on some quite remarkable
qualitative differences compared to the Einstein gravity. We will
consider non-null intersecting hypersurfaces.

The bulk field equation in terms of tensors takes the form:
\begin{gather}\label{Lovelock_tensors}
\sum_{n} \beta_n\ H^\mu_{\ \nu} = T^{\mu}_{{\nu}}\ ,
\\
 H^\mu_{\ \nu}
:= - 2^{-n-1}\ \delta_{\nu\nu_1 \cdots \nu_{2n}}^{\mu\mu_1 \cdots
\mu_{2n}} R^{\nu_1 \nu_2}_{\ \ \ \ \mu_1 \mu_2} \cdots R^{\nu_{2p-1}
\nu_{2n}}_{\ \ \ \ \mu_{2n-1} \mu_{2n}}
\end{gather}
where $H^\mu_{\ \nu}$ is the standard Lovelock tensor one obtains by
varying (\ref{Euler density_tensor}) with respect to the metric.

First note that the variation of the bulk action with respect to the
vielbein gives
\begin{gather}
 \sum_n \beta_n \Omega^{b_1b_2} \wedge \dots \wedge \Omega^{b_{2n-1}
 b_{2n}} e_{a b_1 \dots b_{2n}} = -2T^b_{\ a} e_b.
\end{gather}
Above, it is convenient to define the rescaled constants $\beta_n =
(d-2n)! \alpha_n$. The right hand side is defined so as to agree
with (\ref{Lovelock_tensors}) with $T^{\mu}_\nu = e^\mu_b e^a_\nu
T^b_a$. We also introduced the following
\begin{gather*}
 e_{a_1 \dots a_k} = \frac{1}{(d-k)!}\epsilon_{a_1 \dots a_k
 a_{k+1}\dots a_d} E^{a_{k+1}} \wedge \cdots \wedge E^{a_d}.
\end{gather*}
Note that $e = \sqrt{-g}$ is the volume element in $d$ dimensions.
The variation of this volume element with respect to the vielbein
gives $\delta E^a e_a $. The intrinsic volume element on an
intersection of codimension $p$ is defined as follows: Let $N_j$,
$j=1,\dots,p$ be ortho-normal vectors, i.e. $N_i \cdot
N_j=\eta_{ij}$ or $N_i \cdot N_j=\delta_{ij}$, on the normal space
of the intersection $\{01\cdots p\}$. Let
$E^a=(E^{N_1},\dots,E^{N_p},E^{\mu})$ be the adapted vielbein where
$E^{\mu}$ spans the cotangent space of the intersection. We
introduce the following differential forms:
\begin{equation}\label{intersection volume elements}
\tilde e_{\mu \dots \nu}:=  \prod_{i=1}^p (N_i \cdot N_i)\ e_{a_1
\dots a_{p}\mu \dots \nu}(N_{1})^{a_1} \cdots (N_{p})^{a_p}
\end{equation}
In particular, $\tilde{e}$ is the induced volume element on the
intersection and its variation with respect to the vielbein gives
$\delta E^\mu \tilde{e}_\mu$.

On the simplicial intersection of bulk regions
$\{0\},\{1\},\dots,\{p\}$ the junction conditions read
\begin{gather}\label{general explicit junction}
 ({\cal E}_{01\dots p})_a= -2 (T_{01\dots p})_a^b \tilde e_b
\end{gather}
where $(T_{01\dots p})_a^b$ is the stress-energy tensor living on
the intersection and $({\cal E}_{01\dots p})_a$ is simply the
variation with respect to the vielbein of the surface term ${\cal
L}(\omega_0, \dots, \omega_p, E)$. It is given by:
\begin{gather} \lambda^a \wedge  ({\cal E}_{01\dots p})_a \equiv
\sum_{n=p}^{[d/2]}(d-2n) \alpha_n A_p \ \times \nonumber\\ \times
\int_{s_{0...p}}\hspace{-.2in} d^pt \
i^*f\big((\omega_1-\omega_0)\wedge\!\cdots\wedge
(\omega_{p}-\omega_{0})\wedge \Omega(t)^{\wedge(n-p)}\wedge
\lambda\wedge E^{\wedge(d-2n-1)} \big) \nonumber
\end{gather}
where $\lambda$ is an arbitrary vector-valued 1-form. $i^*$ is the
pullback into the given intersection.

The second fundamental form of the hypersurface $\{ij\}$ embedded in
the bulk region $\{i\}$  is~\cite{Eguchi-80}
\begin{equation}\label{second fundamental def}
\theta^{ab}_{ij}=i^*(\omega_i-\omega_{\|})^{ab}=(N_{ij} \cdot
N_{ij}) (N^a_{ij} K^b_{ij}-N^b_{ij}K^a_{ij})
\end{equation}
where $\omega_{\|}$ is the intrinsic connection in $\{ij\}$ and
$i^*$ pulls the form back into this hypersurface. $N_{ji}=-N_{ij}$
by definition. $N_{ij} \cdot N_{ij}=\pm 1$. The one-form $K^a$
introduced is related to the extrinsic curvature tensor by $K^a :=
K^{a}_{\ b} E^b$. We will use the following convention: $K^a_{ij}$
is the extrinsic curvature of the hypersurface $\{ij\}$ embedded in
the bulk region $\{i\}$ (the first index), and
\begin{equation}
K^{ab}_{ij}=-h^{aa'}D_{a'} N^b_{ij} \: ,\: i>j \quad ; \quad
K^{ab}_{ij}=+h^{aa'}D_{a'} N^b_{ij} \: ,\: i<j.
\end{equation}
Under this convention let us define
\begin{equation}
K^a_{[ij]}:=K^a_{ij}-K^a_{ji}.
\end{equation}
This is the jump of the extrinsic curvature across this
hypersurface. Now consider a product of connection jumps
$\omega_i-\omega_j$ as in (\ref{general explicit junction}) pulled
back into an intersection. $\{ij\}$ is one of the hypersurfaces
involved. As mentioned above the purely tangential components of the
connection are continuous across hypersurfaces. Only the components
$i^*(\omega_i-\omega_j)^{aN_{ij}}$ are non-zero, where $i^*$ is the
pullback into $\{ij\}$. By (\ref{second fundamental def}) we have
that $i^*(\omega_i-\omega_j)^{ab}=(N_{ij} \cdot N_{ij}) (N^a_{ij}
K^b_{[ij]}-N^b_{ij}K^a_{[ij]})$.

After some calculation, using the identity\begin{equation*}\label{}
i^* E^{{\nu}_1...{\nu}_n} \wedge \tilde
e_{{\mu}_1...{\mu}_m}=\frac{m!}{(m-n)!}
\delta^{{\nu}_1}_{[{\mu}_{m-n+1}} \cdots \delta^{{\nu}_n}_{{\mu}_m}
\: \tilde e_{{\mu}_1...{\mu}_{m-n}]}\, ,
\end{equation*}
we find that the junction condition is:
\begin{eqnarray}\label{quite final junction condition}
&& (T_{01\dots p})_{\sigma}^{\tau}=-\sum_{n=p}^{[d/2]} \varsigma_{0
\dots p}\ \beta_n \,2^{p-1} \, \frac{n!}{(n-p)!}\, (2n+1-p)! \times
\\\nonumber && \times \det(M^j_{\:\: i}) \int_{s_{01\dots p}} d^pt
\, (K_{[10]})^{\nu_{1}}_{[\nu_{1}} \cdots
(K_{[p0]})^{\nu_{p}}_{\nu_{p}}\, \Omega(t)^{\nu_{p+1} \dots
\nu_{2n-p}}_{\nu_{p+1}\dots \nu_{2n-p}} \delta^{\tau}_{\sigma]}
\end{eqnarray}
where
\begin{gather*}
 \varsigma_{0
\dots p} := \prod_{i =1}^{p} N_{i0} \cdot N_{i0}  \prod_{j=1}^{p}
N_j \cdot N_j
\end{gather*}
is $+1$ for a space-timelike intersection and $\pm 1$ for a
spacelike intersection depending on the arrangement of the
hypersurfaces which meet there. Also we have defined the matrix
$M^j_{\:\:i}$ by
\begin{equation*}
(N_{i0})^a =\sum_{i=1}^p M^j_{\:\:i} \, N_j^a\, .
\end{equation*}
Compact notation of the form $\Omega^{\nu_{1} \dots
\nu_{2k}}_{\nu_{1}\dots \nu_{2k}}$ standing for
$\Omega^{\nu_{1}\nu_{2}}_{[\nu_{1}\nu_{2}} \dots
\Omega^{\nu_{2k-1}\nu_{2k}}_{\nu_{2k-1}\nu_{2k}]}$ will be used for
convenience.

 The curvature $\Omega(t)$ can be expressed in terms of the
curvatures and the connections of the individual regions by the
symmetrical formula (\ref{symOmegat}). One may write the purely
tangential components of this curvature in terms of curvature of the
bulk regions and quadratic terms in the extrinsic curvatures.
\begin{gather}
 \Omega(t)^{\mu\nu}_{\ \ \kappa \sigma} =
 \frac{1}{2}\left( \sum_i t^i (R_i)^{\mu\nu}_{\ \ \kappa\sigma}
 + \sum_{i>j} t^i t^j (N_{ij}\cdot N_{ij})( K^{\ \ \mu}_{[ij]\, \kappa}
 K^{\ \ \nu}_{[ij] \, \sigma}-
  K^{\ \ \mu}_{[ij]\, \sigma}
 K^{\ \ \nu}_{[ij] \, \kappa} )\right)\label{Omega_t_tensors}
\end{gather}
where $(R_i)^{\mu\nu}_{\ \ \kappa\sigma}$ are the tangential
components of the Riemann tensor in the bulk region $\{i\}$, pulled
back onto the intersection.


Equations (\ref{quite final junction condition}) and
(\ref{Omega_t_tensors}) give the building blocks for writing the
junction conditions for any non-null intersection. The most explicit
formula for the junction conditions involves integrating over the
simplex. This can always be done using the integrals
\begin{equation*}\label{}
\int_{s_{01\dots p}} d^pt\ t_0^{n_0} \dots t_p^{n_p}=\frac{n_0!
\dots n_p!}{(p+\sum_{i=0}^p n_i)!}
\end{equation*}
 but the final
expression can be rather involved.

\begin{center} *** \end{center}
 Let us now look at some of the important differences between
Einstein gravity and higher order Lovelock gravity. Consider a
single hypersurface separating bulk regions labelled by 0 and 1.
Unlike Einstein gravity (Israel junction condition) a vanishing
(non-null) hypersurface's energy tensor
\begin{equation}
T_{01}=0
\end{equation}
does not imply continuity of the connection (that is, zero jump of
the extrinsic curvature) in Lovelock gravity: the relevant
Lagrangian involves polynomial terms of the extrinsic curvatures and
of the intrinsic curvature of the hypersurface and the various terms
may well cancel each other. In general, one cannot deduce an
explicit expression for the jump of the extrinsic curvature from
$T_{ij}$ as one can do in Einstein case.

Let us see this difference explicitly. First, consider Einstein
gravity with cosmological constant, that is, only $\beta_1$ and
$\beta_0$ are non-zero. For the bulk we have
\begin{equation}
G^b_a=\frac{1}{\beta_1} T^b_a+\frac{\beta_0}{2\beta_1} \delta^b_a
\end{equation}
where $G^b_a$ is the Einstein's tensor. The junction conditions for
the hypersurface read
\begin{equation}
(K_{[10]})_{\sigma}^{\tau}-K_{[10]}\,
\delta_{\sigma}^{\tau}=\frac{1}{\beta_1} (T_{01})_{\sigma}^{\tau}
\end{equation}
As is well known, $T_{01}=0$ if and only if
$(K_{[10]})^{\tau}_{\sigma}=0$.

 In the general case, the
junction conditions for the hypersurface are
\begin{equation}\label{1expjunc}
\sum_{n=1}^{[d/2]} \beta_n \,(2n)! \, n  \int_0^1 dt \,
(K_{[10]})^{\nu_{2}}_{[\nu_{2}} \, \Omega(t)^{\nu_{3} \dots
\nu_{2n}}_{\nu_{3} \dots \nu_{2n}}
\delta_{\sigma]}^{\tau}=-(T_{01})_{\sigma}^{\tau}
\end{equation}
Vanishing of the energy tensor $T_{01}$ does \emph{not} imply that
the extrinsic curvature is continuous across the hypersurface in
Lovelock gravity.

Let the discontinuity $K_{[10]}$ be infinitesimal. Then the junction
condition reads
\begin{equation}
\sum_{n=1}^{[d/2]} \beta_n \, 2^{-n+1} (2n)! \, n \,
(K_{[10]})^{\nu_{2}}_{[\nu_{2}} \, R^{\nu_{3} \dots
\nu_{2n}}_{\nu_{3} \dots \nu_{2n}}
\delta_{\sigma]}^{\tau}=-(T_{01})_{\sigma}^{\tau}
\end{equation}
to first order in $K_{[10]}$. Then $T_{01}=0$ for an arbitrary
infinitesimal $K_{[10]}$ if and only if the quantity
\begin{equation}
\sum_{n=1}^{[d/2]} \beta_n \, 2^{-n+1}\,(2n)! \, n  \
\delta^{\nu}_{[\mu} \delta^{\tau}_{\ \sigma} R^{\nu_{3} \dots
\nu_{2n}}_{\nu_{3} \dots \nu_{2n}]}
\end{equation}
vanishes. Inversely, we may think of it as a matrix $M_{IJ} \equiv
M_{(\mu\nu)(\sigma\tau)}$. If the Cauchy data evolve under the
condition that $\det M_{IJ} \neq 0$, then (infinitesimal)
$K_{[10]} = 0$ if and only if $T_{01} = 0$,  that is, $K_{[10]}
\neq 0$ if and only if $T_{01} \neq 0$ so that the extrinsic
curvature can't get a discontinuity across a hypersurface without
a $T_{01}$. This is Choquet-Bruhat's condition for a well posed
initial value problem in Lovelock
gravity~\cite{Choquet-Bruhat:1988dw}. When this determinant
becomes zero, there is a breakdown of predictability in the
theory~\cite{Teitelboim,Deruelle}, which is a key problem to be
addressed if Lovelock gravity is to be regarded as a physical
theory of gravity.
\\

The next more complicated thing than the hypersurface is the
codimension 2 intersection. The simplest case comes from the
Gauss-Bonnet term and reads
\begin{align}\label{}
\qquad &  2 X_{\sigma}^{\tau}-\delta_{\sigma}^{\tau} X_{\rho}^{\rho}
=(T_{012})_{\sigma}^{\tau}
\\ & X_{\sigma}^{\tau} \equiv 4\beta_2 \det(M^j_{\ i})\, \big(
(K_{[10]})^{\rho}_{\rho}
(K_{[20]})_{\sigma}^{\tau}+(K_{[20]})^{\rho}_{\rho}
(K_{[10]})_{\sigma}^{\tau}-2(K_{[10]})^{\rho}_{\sigma}(K_{[20]})^{\tau}_{\rho}
\big) \nonumber
\end{align}
Note first that the energy tensor $T_{012}$ on the intersection
vanishes if and only if the matrix $X_{\sigma}^{\tau}$ vanishes.
This does not imply that the jumps $K_{[10]}$ and $K_{[20]}$ vanish
too. On the other hand, intersecting or colliding shells of matter
will in general produce a non-zero energy tensor on the hypersurface
where their spacetime trajectories meet. This is unavoidable however
small $\beta_2$ might be.

The most obvious physical implication coming out of this work and
the previous one~\cite{Gravanis}, is that intersections in
Lovelock gravity involves hypersurfaces  of various
codimensionalities carrying non-zero energy tensors.In particular
one can consider a collision of shells, that is, a spacelike
intersection $C$ of timelike hypersurfaces, with a total of $m$
ingoing and outgoing shells. There could exist a non-zero stress
tensor on $C$, $T_C$. (The appearance of a stress tensor on the
spacelike surface would have to be due to some exotic kind of
matter, in conflict with the dominant energy
condition~\cite{Hawking-73} as mentioned already
in~\cite{Gravanis}.) An implication is that we may have $m$
outgoing or $m$ ingoing shells i.e. the shells of matter may all
originate from or disappear into $C$. This also is forbidden in
Einstein gravity by the conservation of energy for positive energy
densities, but perfectly possible here because of $T_C$ (see
section 3.2 of~\cite{Gravanis} for the energy exchange relations).
The reason is that the energy exchange relations involve also
extrinsic curvatures and $C$ acts as \emph{source} which can emit
or absorb all $m$ timelike trajectories. The dominant energy
condition is respected in Lovelock gravity collisions if the bulk
geometries are constraints by the condition $T_C=0$.

We have mentioned that hypersurfaces with no energy tensor are
possible in Lovelock gravity. That is a discontinuity of the
connection can be self-supported. We showed in ref. \cite{soon}
that one may have a bulk AdS spacetime vacuum with such
discontinuities. More complicated cases don't seem impossible. If
perfect homogeneity is given up, we have an interesting kind of
vacuum in this gravity.

Continuing with matters of vacuum, thin shells and their
gravitational effects appear when separating the phases of false and
true vacuum in false vacuum decay in the presence of
gravity~\cite{Coleman:1980aw}. This happens when there are more one
(local) minima of the energy of a system and not all of them have
the same value. A (only) local minimum state, false vacuum, decays
by formation of bubbles of true vacuum which grow very fast and
eventually collide. The bubble effects play an important role in the
inflationary evolution model of the early
universe~\cite{Guth:1980zm}, the implications of collisions have
been studied in~\cite{Hawking:1982ga}. In a `universe' with more
than four dimensions at those times or in general, the collisions of
the bubbles have additional effects as we have learned here: at the
spacelike hypersurface of collision there will in general live a non
zero stress tensor. That is, an instanton-like configuration of non
topological nature. These may have interesting implications, they
are though beyond the scope of this paper.

\section{Conclusion}

The theory of General Relativity (GR) admits singular sources whose
stress-energy tensor has support on a hypersurface. In general, an
arbitrary collection of such objects should intersect. We have shown
that in gravities including dimensionally continued Euler densities
with up to $n$ factors of the Riemann tensor, not only are such
hypersurface sources well defined but that there is a possibility of
sources of codimension up to $n$ at the intersections. This becomes
possible because the equations of motion are not linear in curvature
and of the fully anti-symmetric way the Riemann tensors are
contracted. To give an example, imagine we have two hypersurfaces
crossing each other, one at $x=0$ and the other where $y=0$. Then
some components of the curvature will have a $\delta(x)$ singularity
and some other will have a $\delta(y)$ one. If the theory of gravity
includes a Gauss-Bonnet term, the energy tensor will have a term of
the form $\delta(x)\delta(y)$ i.e. a codimension two matter
distribution localized at $x=y=0$. In Einstein theory if the
curvature is of that form one could not have a codimension two
matter. Schematically,
\begin{gather*}
f(\Omega \wedge E^{\wedge(d-2)})\approx A\delta(x) + B\delta(y),\\
f(\Omega^{\wedge2}\wedge E^{\wedge(d-4)}) \approx C\delta(x,y).
\end{gather*}
A $\delta(x,y)$ could be produced in Einstein gravity only if the
curvature itself had such a singularity, which implies a conical
singularity.
\\

The $n=1$ (GR) junction conditions give a 1-1 correspondence between
the discontinuity of the connection and the energy-momentum tensor
on a hypersurface. For the Lovelock theory with higher $n$ terms,
things are more complicated: the energy-momentum tensor is a
polynomial in the curvature and discontinuity of the connection. For
a singular energy-momentum tensor to be supported, there must be a
discontinuity, but the converse does not apply. It is possible for
the energy-momentum tensor to vanish even if there is a
discontinuity.
\\

The metric describing an intersection of hypersurfaces which in GR
has no localized matter at the intersection, will generally produce
localized matter due to the non-trivial junction conditions for the
higher order Lovelock terms. `Intersection' is a general term and
includes the case where the intersection hypersurface is space-like
where we have a collision. Then, if we demand no localized
space-like matter there will be a constraint on the geometry. The
constraint will be of order $\alpha_2$, the coefficient of the
quadratic Lovelock term. Thus, the higher order Lovelock terms place
additional constraints on the way that singular matter sources can
interact with each other. This qualitative difference is well
illustrated by a planar intersection in AdS space~\cite{soon}.
\\

Expressions like $P\!\left([d_{(t)} A(t) + F(t)]^{\wedge n}\right)$,
descended from a Characteristic Class $P\!\left(F^{\wedge
n}\right)$, are already known in the mathematics
literature~\cite{Gabrielov} and in the context of anomalies in gauge
theory or gravity~\cite{Guo}. The homotopy operator
(\ref{compisCartan}) has appeared in ref.~\cite{Gelfand2}. We have
shown that these geometrical methods are very useful when studying
intersections of hypersurfaces in gravity theories.
\\\\

\begin{centering} {\bf
Acknowledgements}\\\end{centering}\vspace{.15in}

S.W. would like to thank CERN Theory division, where this work was
done, for their hospitality. S.W. was funded by European Union,
contract HPRN-CT-2000-00152.

\appendix
\numberwithin{equation}{section}

\section{Proof of $d_F\eta =0$.}

Recall the definition of $\Omega(t)$ and also the Jacobi
identity\cite{Chern}:
\begin{gather}
\Omega(t)=d_{(x)}\omega(t)+\frac{1}{2}[\omega(t),\omega(t)]\\
[[\omega,\omega],\omega]=0
\end{gather}
From these one can easily find the following identities.
\begin{gather}
d_{(x)} [\omega(t),\omega(t)] = 2[\Omega,\omega(t)]\label{A4}\\
d_{(t)} [\omega(t),\omega(t)] = 2[d_{(t)}\omega(t),\omega(t)]\\
\label{kindofC}d_{(t)}\omega(t)+\Omega(t) = d_F\omega(t)+\frac{1}{2}
[\omega(t),\omega(t)]\end{gather} Also, from (\ref{A4}), we get the
Bianchi identity for $\omega(t)$:
\begin{gather}\label{Bianchi}
D(t)\Omega(t) = 0
\end{gather}
Like $\omega_0$, $\omega(t)$ is a connection and so the invariance
property of $f$ implies, for 2-forms $\psi$:
\begin{gather}\label{Invariance}
\sum_i f(\psi_1\wedge...[\omega(t),\psi_i]...\wedge\psi_n) = 0
\end{gather}
Combining $(A.4-6)$:
\begin{gather}\label{DFeta}
d_F \big(d_{(t)}\omega(t)+\Omega(t)\big)=
[d_{(t)}\omega(t)+\Omega(t),\omega(t)]
\end{gather}
which is equivalent to (\ref{Fbianchi}) and so our Proposition 2
follows by the invariance property of the Polynomial
(\ref{Invariance}).
\begin{gather}\label{dfvanish}
d_F f\Big(\big[d_{(t)}\omega(t)+\Omega(t)\big]^{\wedge n}\Big)= 0
\end{gather}

Let us expand the polynomial:
\begin{align}\label{expand}
\big[d_{(t)}\omega(t)+\Omega(t)\big]^{\wedge n} =&\sum_{l=0}^{n}\
^nC_l \left(\sum_{\alpha=1}^pd_{(t)}\omega(t)\right)^{\wedge l}
\wedge\Omega(t)^{\wedge (n-l)}
\\\nonumber =&\sum_{l=0}^n(-1)^{l(l-1)/2}
\ ^nC_l \ dt^{i_1}\!\wedge \cdot\cdot\cdot\wedge\!
dt^{i_l}\wedge\\\nonumber & \hspace{1.5in}\wedge
d_{(t)}\omega\!(t)_{i_1}\wedge\cdot\cdot\cdot \wedge
d_{(t)}\omega\!(t)_{i_l} \wedge\Omega(t)^{\wedge (n-l)}.
\end{align}
The first term in the expansion evaluated at the 0-simplex $s_{i}$
is just the Euler density $(\ref{POmega}$) in the interior of the
region $i$. Thus (\ref{dfvanish}), combined with (\ref{equiv})
completes the proof by induction of the second proposition. As a
consistency check, we can see that the terms in this expansion
reproduce the form of $(\ref{Lagrange})$.

\section{$W$-space and topology}\label{topappendix}

The closure of $\eta$ in $F$ but also means that it obeys the same
transgression formula as the invariant polynomial we started with,
only now on $F$. Under continuous variation $\omega(t)\rightarrow
\omega'(t)$,
\begin{gather} \label{transeta}
P(\Omega_F) -P(\Omega'_F) =d_F T\!P(\omega(t),\omega'(t)).
\end{gather}

Let us for now assume that $M$ is compact. We define a covering of
open sets on $W$ by the open sets on $M$. Choose a covering of $M$.
For every open $O_i \subset M$ define the set ${\cal O}_i \subset W$
as the set $\{(t,x) \in W| \; x \in O_i\}$. Clearly ${\cal O}_i$'s
cover $W$ and are open sets\footnote{One may face problems only if
an infinite number of intersections is considered, locally. Apart
from that, the t-excursions add no accumulation points, that is
boundary points, to the sets $O_i$.}, endowing $W$ with a manifold
structure. This gives $W$ the topology of $M$. For a partition of
unity $f_i$ of $M$ we define the partition of unity of $W$ simply by
$f_i(t,x)=f_i(x)$. Then, by the invariance (\ref{invP}),
(\ref{etaisEuler}) is meaningful over $W$ associated with a
topologically non-trivial $M$ just as $\int_M {\cal L}(\omega)$ is
meaningful over $M$.

The shape of $W$ is interesting. Every $d-1$ dimensional surface is
thickened in the t-direction by a 1-dimensional simplex; These meet
at a $d-2$ surface in M which looks like a triangular prism in $W$
(fig. \ref{simp} (c)), etc. We know that the equality holds:
\begin{gather} \label{eulerno1}
\int_M {\cal L}(\omega) = \int_{W} \eta(\omega(t))\quad (\propto
{\rm Euler~no.})
\end{gather}
All that we did in Section \ref{Geomsection} amounts to expanding
both sides via (\ref{rewrite}) and (\ref{expand}), and equating the
terms. Given that $M$ and $W$ have the same topology (and Euler
number) we can say that $\eta(\omega(t))$ is the Euler density of
$W$.

If we calculate $\int_M {\cal L}(\omega)$ with a different $C^0$
metric \footnote{The argument of ${\cal L}$,  $\omega$, is unrelated
to $\omega(t)$; they are just both assumed to produce the same Euler
number. As we explained in \cite{Gravanis}, $\omega$ is associated
with a smooth metric, say $C^2$. Our formulas give the wanted
independence from the connection, of certain topological quantities,
when discontinuities are allowed.}(with discontinuities of the
connection at intersecting hypersurfaces) described by an
$\omega'(t)$, we have along with (\ref{eulerno1}) the relation
\begin{gather} \label{eulerno2}
\int_M {\cal L}(\omega) = \int_{W} \eta(\omega'(t))
\end{gather}
Then (\ref{transeta}) tells us that
\begin{equation}
\int_W d_F TP(\omega(t),\omega'(t))=0
\end{equation}
This is true for quite arbitrary $\omega(t),\omega'(t)$ so we must
have
\begin{equation} \label{Wcycle}
\partial_F W=0.
\end{equation}
\\\\
{\bf Proposition (B1).} Define $\pi: W \to M$, $\pi (x,t) =x$. Then
$\pi(\partial_FW) \subset
\partial M$ if and only if:
\begin{equation*}
\partial\{i_0...i_p\}=\sum_{i_{p+1}} \{i_0..i_pi_{p+1}\} +
\{i_0...i_p\}\cap \partial M.
\end{equation*}
In particular $\partial_F W =0$ if $\partial M=0$. These relations
can be taken as the definition of the simplicial intersections,
which we used in ref.~\cite{Gravanis}.
\\\\
{\bf Proof:} When each $\omega_i, E_i$ are chosen to be that of each
bulk region i, then:
\begin{equation}\label{Walanerve}
W=\sum_{p=0}^{h}\sum_{i_0\ldots i_p} \frac{1}{(p+1)!}\, s_{i_0\ldots
i_p} \times \{{i_0\ldots i_p}\} \subset F
\end{equation}
where $\{i_0...i_p\} \subset M $ is a codimension $p$ sub-manifold
and as a point-set corresponds to the codimension $p$ simplicial
intersection. $h$ is the codimension of the highest codimension
intersection present. Note that it is sufficient to take
$\{i_0...i_p\}$ fully anti-symmetric.

Then,
$$\partial_F W=\sum_{p=0}^{h}\sum_{i_0\ldots i_p}
\frac{1}{(p+1)!} \left\{ \partial s_{i_0\ldots i_p} \times
\{{i_0\ldots i_p}\}+(-1)^p  s_{i_0\ldots i_p} \times \partial
\{{i_0\ldots i_p}\} \right\},$$ where $\partial_{(x)} \;
s_p=(-1)^ps_p \;
\partial_{(x)}$ was used, and $\partial_F=\partial_{(x)}+\partial_{(t)}$.
For the first term we have
\begin{equation}
\begin{split}
&\sum_{p=0}^{h}\sum_{i_0\ldots i_p} \frac{1}{(p+1)!} \sum_{r=0}^{p}
(-1)^r  s_{i_0\ldots \hat{i}_r \ldots  i_p} \times \{{i_0\ldots
i_p}\}=\\
&\sum_{p=1}^{h}\sum_{i_0\ldots i_p} \frac{1}{(p+1)!} (p+1)(-1)^p
s_{i_0\ldots i_{p-1}} \times \{{i_0\ldots
i_p}\}= \\
&\sum_{p=0}^{h-1}\sum_{i_0\ldots i_pi_{p+1}} \frac{1}{(p+1)!}
(-1)^{p+1} s_{i_0\ldots i_{p}} \times \{{i_0\ldots i_pi_{p+1}}\}
\end{split}
\end{equation}
so combining with the second term we have
\begin{eqnarray} \label{findW}
&& \partial_F W=\sum_{p=0}^{h-1} \sum_{i_0...i_p} (-1)^p
\frac{1}{(p+1)!} s_{i_0...i_p} \times \{ -\sum_{i_{p+1}}
\{i_0...i_pi_{p+1}\}+
\partial \{i_0...i_p\}
\} + \\
&& \qquad +(-1)^h s_{01...h} \times \partial \{01...h\} \nonumber
\end{eqnarray}
For the highest codimension surface one has
\begin{equation}
\partial \{01 \ldots h \}=\{01...h\} \cap \partial M.
\end{equation}
thus we get Proposition (B1). If we ignore boundary terms on
$\partial M$, we may justly ignore boundary terms on $\partial_F W$.

\section{Invariance of $\eta$.}


In this Appendix it is shown that $\eta$ form is invariant under
gauge transformations:
\begin{gather}\label{invP}
\eta (\omega(t)) = \eta (\omega(t)_{(g)})
\end{gather}

Under the change $\omega_{i} \to {\omega_{i}}_{(g)}$ of the
connection of every region $\{i\}$ with
\begin{equation}
{\omega_{i}}_{(g)}=g^{-1} \omega_{i} g+ g^{-1} d_{(x)} g
\end{equation}
The interpolating connection $\omega(t)=\sum_{i=0}^p t^{i}
\omega_{i}$ (with $\sum_{i=0}^p t^{i}=1$) changes as $\omega(t) \to
\omega(t)_{(g)}$ with
\begin{equation} \label{gtransconn}
\omega(t)_{(g)}=g^{-1} \omega(t) g+ g^{-1} d_{(x)} g
\end{equation}
Then both
\begin{equation}
d_{(t)} \omega(t)= \sum_{i=1}^p dt^i (\omega_i-\omega_0) \quad ,
\quad \Omega(t)=d_{(x)} \omega(t)+ \frac{1}{2}[\omega(t),\omega(t)]
\end{equation}
obviously transform as
\begin{equation}
d_{(t)}\omega(t)_{(g)}=g^{-1} d_{(t)} \omega(t) g \quad , \quad
\Omega(t)_{(g)}=g^{-1} \Omega(t) g
\end{equation}
so $\Omega_F$ itself changes to
\begin{equation} \label{gtranscurv}
({\Omega_F})_{(g)}=g^{-1} \Omega_F g
\end{equation}
and the invariant polynomial $P$ gives us the wanted invariance
relation (\ref{invP}).

 Actually, the abstraction of $\Omega_F$ as
a curvature associated to $\omega(t)$ and the derivative operator
$d_F$ helps us again to prove things easily. By its very definition
(which let us repeat)
\begin{equation}
\Omega_F=d_F \omega(t) +\frac{1}{2} [\omega(t),\omega(t)] \nonumber
\end{equation}
we see that under $\omega(t) \to \omega(t)_{(g)}$ with
\begin{equation}
\omega(t)_{(g)}=g^{-1} \omega(t) g + g^{-1} d_F g
\end{equation}
we have immediately the covariant transformation (\ref{gtranscurv}).
But $d_Fg=d_{(x)}g$
so 
we proved again (\ref{invP}).

\end{document}